\documentstyle[proceedings,epsf]{crckapb}

\begin{opening}
\title{THE ROLE OF BINARIES IN THE DYNAMICAL EVOLUTION OF THE CORE OF
A GLOBULAR CLUSTER}

\author{Piet Hut}
\institute{Institute for Advanced Study, Princeton, NJ 08540, U.S.A.}

\end{opening}

\runningtitle{The Role of Binaries}

\begin{document}

\def\half{{\scriptstyle {1 \over 2}}}
\def\ie{{\it {\frenchspacing i.{\thinspace}e. }}}
\def\eg{{\frenchspacing e.{\thinspace}g. }}
\def\cf{{\frenchspacing\it cf. }}
\def\et{{\frenchspacing\it et al.}}
\def\simlt{\hbox{ \rlap{\raise 0.425ex\hbox{$<$}}\lower 0.65ex\hbox{$\sim$} }}
\def\simgt{\hbox{ \rlap{\raise 0.425ex\hbox{$>$}}\lower 0.65ex\hbox{$\sim$} }}
\def\solar{\odot}
\def\msun{{\rm M}_\odot}
\def\rsun{{\rm R}_\odot}
\def\pc{{\rm pc}}
\def\Myr{{\rm Myr}}
\def\Gyr{{\rm Gyr}}
\def\Rf{\parindent=0pt\medskip\hangindent=3pc\hangafter=1}

\begin{abstract}

The size of the core is one of the main diagnostics of the
evolutionary state of a globular cluster.  Much has been learned over
the last few years about the behavior of the core radius during and
after core collapse, under a variety of different conditions related
to the presence or absence of large numbers of binaries.  An overview
is presented of the basic physical principles that can be used to
estimate the core radius.  Four different situations are discussed,
and expressions are presented for the ratio $r_c/r_h$ of core radius
to half mass radius.  The regimes are: deep collapse in the absence of
primordial binaries; steady post-collapse evolution after primordial
binaries have been burned up; chaotic post-collapse evolution under
the same conditions; and post-collapse evolution in the presence of
primordial binaries.  In addition, modifications to all of these cases
are indicated for the more realistic situation where effects of the
galactic tidal field are taken into account.

\end{abstract}

\section{Introduction}

Core collapse is the most dramatic phenomenon in the evolution of
globular clusters.  It corresponds to the pre-main-sequence stage in
stellar evolution, when the energy losses from the photosphere are not
yet balanced by nuclear energy generation in the center.  In contrast,
during the post-collapse evolution of a star cluster some form of
central energy source turns on, which makes up for the energy lost
through conduction and evaporation of stars from the central regions
(for a review of different mechanisms, see Goodman 1993).

The details of the halt of core collapse only emerged some dozen years
ago, and were reported in IAU symposium 113 (Goodman \& Hut, 1985).  A
few years later, the plot thickened again through the discovery that a
significant fraction of stars in globular clusters were contained in
binaries.  These primordial binaries modify the picture of core
collapse considerably.  Rather than proceeding down to the miniscule
core and enormous densities required in the earlier single-star
models, core collapse is halted at moderate central densities of order
$\rho_c \sim 10^6 M_\solar\pc^{-3}$, and a typical core radius of $r_c
\sim 0.1 \pc$.  The corresponding core contains a few $\times 10^3
M_\solar$, roughly one percent of the inner half mass of the cluster
enclosed in a radius $r_h \sim 10\pc$ (Goodman \& Hut, 1989).

A few years ago, we have given a detailed review of primordial
binaries in globular clusters, and their effects (Hut \et\ 1992).  I
refer to this paper for many original references, as well as summaries
of the history of this field.  Many of the more recent theoretical
developments concerning the role of binaries in star clusters are
discussed by others in these proceedings, \eg\ in the contributions by
Aarseth, Clarke, Heggie \et, Kiseleva, Leonard, Mardling, McMillan,
Phinney, and Rasio \& Heggie.  In order to avoid too much overlap with
these other papers, I will concentrate here on a single question that
is perhaps most relevant to observations: how to estimate the core
size $r_c$ of a post-collapse cluster.  In four separate situations, I
will describe the basic physics from which the ratio $r_c/r_h$ can be
estimated, where $r_h$ is the half-mass radius.  These are given in
\S\S 2-5 below.  \S6 discusses the consequences of adding a galactic
tidal field, and \S 7 sums up.

\section{Deep Single-Star Core Collapse}

In the late stages of core collapse, the core dynamically decouples
from the bulk of the cluster, in a process called `gravothermal
catastrophe' (Lynden-Bell \& Wood 1968; Spitzer 1987).  The reason for
this instability is the fact that dynamical evolution proceeds on a
relaxation time scale.  The relaxation rate is proportional to the
number of encounters per star, which in turn is proportional to the
density, which is far higher in the core of an evolved cluster than in
the bulk of the cluster.  Therefore, when the core is losing energy by
local `evaporation' of stars into the surrounding regions, as well as
by heat conduction into those regions, the rest of the material has no
time to keep up with the shrinking core.  Collapse will occur,
leading to an infinite density in finite time, unless some other
physical process will switch on.

Let us make a rough estimate of the critical size for the core of a
star cluster, when core collapse is halted.  In the equal-mass
point-mass approximation, heat production is proportional to the rate
of binary formation in three-body encounters, a process that is
proportional to the third power of the density:
\begin{equation}
{\dot E}_+ = C_3 \rho^3.
\end{equation}
Energy losses, caused by two-body relaxation, are proportional to the
number of encounters per unit volume, which in turn is proportional to
the square of the density $\rho$:
\begin{equation}
{\dot E}_- = C_2 \rho^2.
\end{equation}

A natural local choice of physical units is one in which the
gravitational constant $G$, the mass of a single star $m$, and the
root-mean-square velocity of the single stars $v$ are all set equal to
unity: $G = m = v = 1$. Since the core is close to being
self-gravitating, we can use the virial theorem: 
\begin{equation}
N_c(\frac{1}{2}mv^2) = \frac{G(mN_c)^2}{4r_c},
\end{equation}
where $N_c$ is the number of particles in the core and $r_c$ the core size.
We thus find $N_c = 2r_c$, and $\rho = N/((4/3)\pi r_c^3) \simeq 2N^{-2}$.
Requiring now that ${\dot E}_+ = {\dot E}_-$, we find
\begin{equation}
N_c = \sqrt{\frac{2C_3}{C_2}} = \sqrt{\frac{20}{0.003}} \simeq 80,
\end{equation}
where the estimates for the numerical values for the constants are
taken from Hut \& Inagaki (1985).  A similar estimate leading to $N_c
\simeq 50$ was made by Goodman (1984).  It is clear that deep core
collapse leads to the momentary appearance of a tiny core with less
than one hundred stars, with $r_c \ll 0.01$pc corresponding to a huge
density, of order $10^9M_\solar\pc^{-3}$, for typical globular cluster
parameters.

\section{Single-Star Post-Collapse Evolution: Smooth Core Reexpansion}

Long after core collapse, on a time scale of several half-mass
relaxation times, the outer regions of the cluster are finally able to
catch up with what has happened in the center.  For small to moderate
numbers of stars ($N \simlt 10^4$; Goodman 1987), the core is able to
adjust itself to the demands of these outer regions.  Given the rate
of energy loss from these regions, the core will find an equilibrium
size and corresponding density such that it will form new binaries at
the correct rate to make up for the outer energy losses.

Goodman (1984) has made a detailed analysis of this situation, and
derived a relation $N_c \propto N^{1/3}$, which implies $r_c/r_h
\propto N^{-2/3}$.  For realistic globular cluster parameters, he
found $N_c \sim 3\times10^2$, six times larger than the deep collapse
figure of $N_c \sim 50$ that he found.  The reason that the core
radius becomes about six times smaller during deep collapse, compared
to subsequent smooth expansion, is that the latter takes place in
near-isothermal equilibrium.  As a result, the conduction rate of
energy through two-body relaxation is much lower than during the
original collapse, which leaves a more substantial temperature
gradient in its wake.  Lowering the conduction rate is equivalent to
lowering the value $C_2$, which has the effect of increasing $N_c$,
according to eq. 4.

\section{Single-Star Post-Collapse Evolution: Chaotic Core Reexpansion}

For large numbers of stars ($N \gg 10^4$), post-collapse expansion of
the core of a star cluster does not proceed smoothly.  As discovered
by Sugimoto and Bettwieser (1983; Bettwieser and Sugimoto 1984),
chaotic fluctuations occur in the size of the core radius.  These can
be explained as a consequence of the gravothermal instability, and
were therefore called `gravothermal oscillations'.

The underlying physical mechanism can be characterized as follows. For
a large number of stars in the system, the inner relaxation timescale
is much larger than the half-mass relaxation timescale, which
determines the overall rate of expansion. Therefore, the inner regions
have the tendency to evolve on a timescale much smaller than the bulk
expansion timescale. As a result, the inner regions tend to get
impatient, and a small fluctuation can trigger a local re-collapse,
followed by a local re-expansion. The larger the number of stars, the
more the central and outer timescales are decoupled, and the more
chaotic the oscillations become.  The dynamical behavior of these
oscillations can be shown to be characterized by a low-dimensional
chaotic attractor (Breeden \et\ 1990; Cohn \et\ 1991; Breeden \& Cohn
1995).  The gravothermal character of the core oscillations was
confirmed explicitly by Goodman (1987), who performed a linear
stability analysis of a new regular self-similar model for
post-collapse evolution, and classified the different modes of
behavior according to the type of linear instability they exhibit.

Like gravothermal collapse, gravothermal oscillations appear to be a
ubiquitous phenomenon, at least in the models which treat the stars
and all physical processes as continuous quantities.  Inagaki (1986)
and McMillan (1986, 1989) have expressed doubts as to whether the
oscillations persist in real clusters, where the stars and the
physical processes are discrete, and statistical fluctuations may be
large.  This issue has now been resolved by direct N-body simulations
of systems containing $>10^4$ stars, and the conclusion is that
oscillations do occur, and indeed have a clearly gravothermal nature
(Makino, this volume).

Although the details of the gravothermal oscillations are strongly
dependent on the total number of stars, as well as their mass
spectrum, the maximum value of the core radius is relatively
insensitive to those details.  Since the evolution slows down most
around the time of maximum expansion, observations of a cluster core
will find the core to be near maximum expansion in the overwhelming
majority of cases (just as a binary star in a very eccentric orbit
will almost always have a separation close to twice the semimajor axis).
Typical values, from Fokker-Planck calculations including a mass
spectrum, are $r_c/r_h \sim 10^{-2}$ (Murphy \et\ 1990).

\section{Post-Collapse Evolution with Primordial Binaries}

\def\fb{{\ifmmode{f_B}\else{$f_B$}\fi}}
\def\fbcc{{\ifmmode{f_{Bc}}\else{$f_{Bc}$}\fi}}
\def\fbc{{\ifmmode{f_{B,crit}}\else{$f_{B,crit}$}\fi}}
\def\fbcl{{\ifmmode{f_B<f_{B,crit}}\else{$f_B<f_{B,crit}$}\fi}}
\def\fbcg{{\ifmmode{f_B>f_{B,crit}}\else{$f_B>f_{B,crit}$}\fi}}

Among globular cluster stars, a fair fraction are born as members of
binaries.  The fraction may not be as large as that in the galactic
disk, but it is large enough to have a significant dynamical effect
(see Hut \et\ 1992 for a detailed review and references).  The overall
cluster binary population is conveniently parametrized by the binary
fraction \fb, defined as the number of ``objects'' in a cluster that
are actually binaries (so, if binary components are representative of
the cluster as a whole, the binary mass fraction is
$\sim2\fb/[1+\fb]$).  Because of the presence of many observational
selection effects, this quantity is not known accurately, but it
probably lies in the range 3--30\%, and a value of $f_B\sim10\%$ is
widely taken to be ``typical.''

The presence of primordial binaries radically changes the picture
sketched above, in which clusters undergo deep core collapse, before
re-expanding, with or without core oscillations.  As was pointed out
by Goodman \& Hut (1989), gravitational `burning' of binaries will
cause the core collapse to be halted at a far larger core radius of
$r_c \approx 0.02 r_h$, compared to $r_c \sim 10^{-4} r_h$ for deep
core collapse.

This phenomenon has its analogue in stellar evolution, where a
protostar does not land directly on the hydrogen main sequence, but
instead spends a brief time on the deuterium main sequence.  While
primordial deuterium is being burned in the center of the star, its
radius remains significantly larger than its eventual value.  The
reason is the high efficiency of deuterium burning at relatively low
temperature, compared to hydrogen burning.  As a result, central heat
production and surface heat loss are balanced at a lower central
temperature, and hence larger stellar radius.

Similarly, the high efficiency of gravitational energy extraction from
an existing binary population, compared to binary production from
single stars only, allows the cluster core to remain relatively large,
until the primordial binaries have been depleted.  Numerical
conformation of the semi-quantitative estimate given above, for
$N$-body simulations with $N \le 2\times 10^3$ (McMillan \et\ 1990,
1991; McMillan and Hut 1994) showed that for $N=(1\sim2)\times10^3$,
the core radius $r_c \approx (0.10 \sim 0.15)r_h$.  This is larger
than the value predicted by Goodman and Hut (for this $N$ range, their
eq. (6) would give $r_c \approx 0.05 r_h$).  Most likely, this
discrepancy is caused by the low $N$ values used in direct $N$-body
simulations so far; it is not clear whether values of $N\sim 10^3$,
with total numbers of core stars $N_c < 10^2$, are high enough to
reach the asymptotic scaling regime.  Fokker-Planck calculations by
Gao \et\ (1991) resulted in values in the range $r_c \approx (0.01
\sim 0.04)r_h$, in broad agreement with the earlier analytical
estimate.  A more definite answer will soon be provided, when we will
run primordial binary simulations on the GRAPE-4, for $N\ge2\times10^4$.

\section{Effects of the Galactic Tidal Field on Post-Collapse Evolution}

\begin{figure}
\begin{center}
\leavevmode
\epsfxsize 11cm
%\epsffile{iau174figs/piet_fig1.ps}
\epsffile{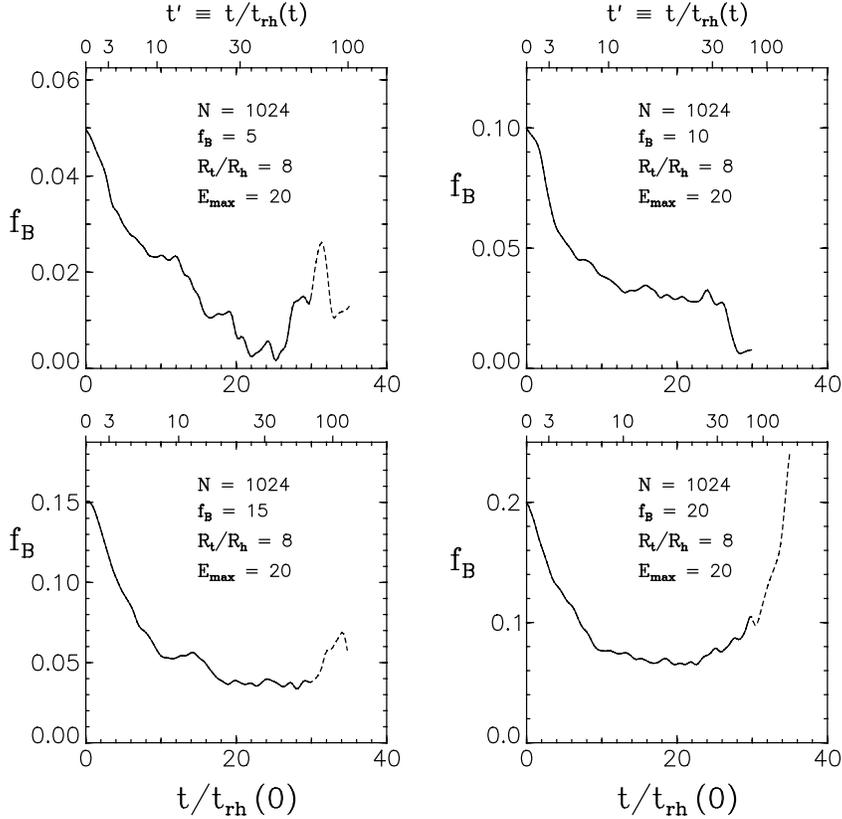}
\caption{Binary fraction \fb\ as a function of time $t$, in units of
the original half-mass relaxation time.  $N$ denotes the number of
particles in the simulation, $R_t/R_h$ the ratio between tidal radius
and half-mass radius, and $E_{max}$ the maximum value in the spectrum
of initial binding energies of the primordial binaries.  The dashed
portion of the lines indicate where the total number of stars has
dropped to such a low value that low-number statistics make the result
unreliable.}
\end{center}
\end{figure} 

Let us now take into account the fact that globular clusters do not
evolve in isolation, but are tidally limited by the presence of the
gravitational field of the galaxy.  This means that the average
density of the cluster is forced to remain at a fixed value,
comparable to the average density of the matter inside its orbit (for
a circular orbit; the situation is qualitatively similar for orbits of
moderate eccentricity).  As a result, $r_h$ is forced to shrink upon
mass loss through evaporation of cluster stars, in stark contrast to
the evolution of isolated clusters, for which $r_h$ grows steadily
after core collapse.

\begin{figure}
\begin{center}
\leavevmode
\epsfxsize 11cm
%\epsffile{iau174figs/piet_fig2.ps}
\epsffile{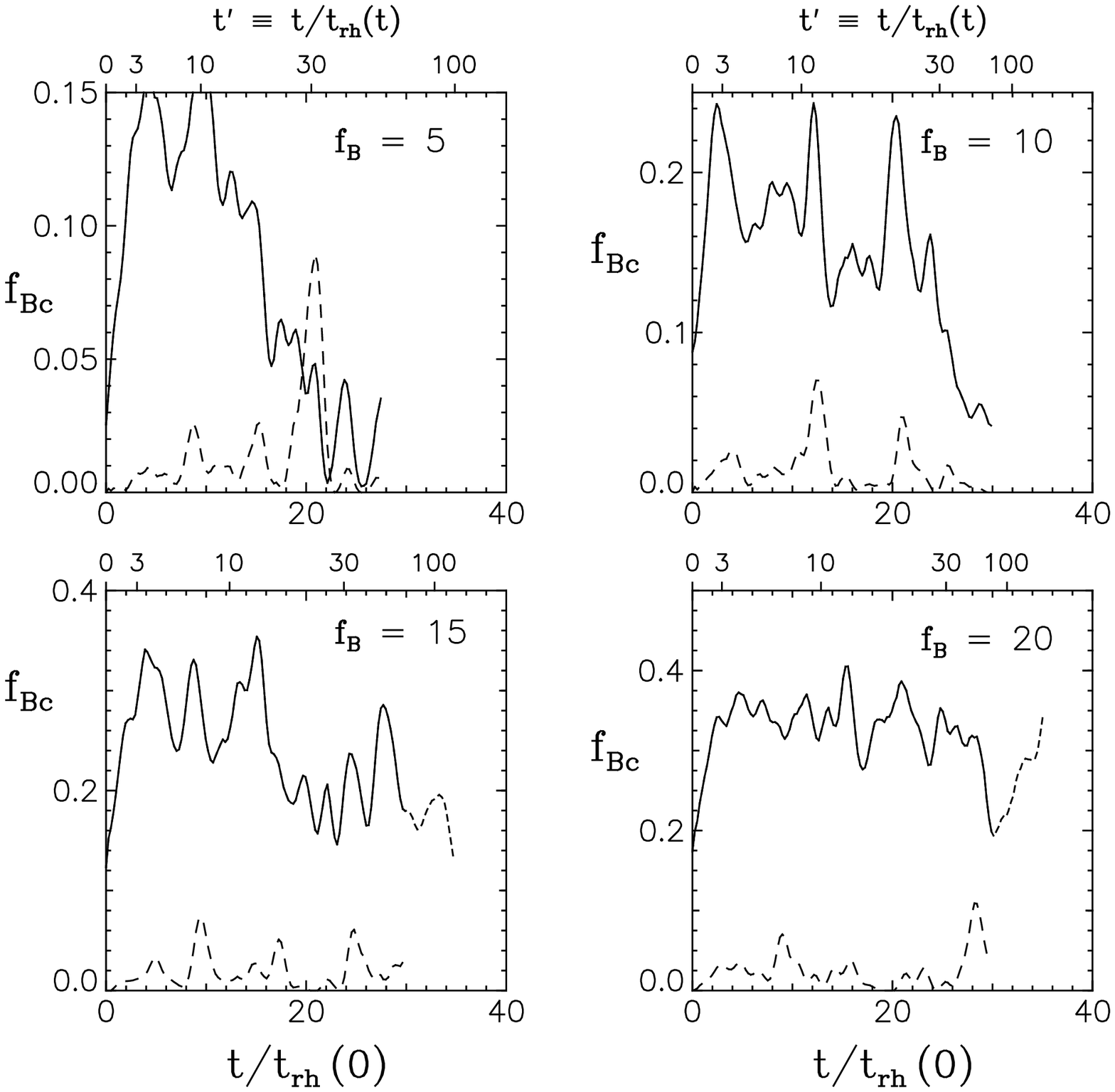}
\caption{Binary fraction \fbcc\ in the core, as a function of
time.  \fb\ denotes the initial value of the overall binary fraction.
The long-dashed line indicates the fraction of triple stars.}
\end{center}
\end{figure} 

Since mass segregation tends to concentrate the heavier binaries
towards the core, most of the evaporating stars are single.
Therefore, the depletion of binaries in the core by gravitational
`burning' may or may not be offset by the depletion of single stars
through evaporation from the outer parts of the cluster.  For each
cluster, with a given mass and size, there is a critical binary
fraction \fbc, forming a watershed between these two possibilities.
For \fbcl, primordial binaries will be burned up before the cluster as
a whole dissolves in the tidal background field.  In this case, the
later stages of cluster evolution will show gravothermal oscillations
(if $N$ is large enough).  However, for \fbcg, primordial binaries
will remain present in the core until the end, and the overall binary
fraction \fb\ will in fact increase in the later stages of the
cluster.  Gravothermal oscillations will not occur during any stage of
cluster evolution, in this case.

These two possibilities are illustrated in fig. 1, based on the
simulations reported by McMillan and Hut (1994).  The watershed value
here is $f_{B,crit} \sim 15\%$.  The precise value of \fbc\ shows some
dependence on the cluster parameters, especially on the size of the
tidal radius $r_t$ (for $r_t < 0.8 r_h$, the standard choice in our
simulations, \fbc\ will be smaller, since single star evaporation will
be more rapid).  In practice, \fbc\ may well be lower than estimated
here, if we take tidal shocking processes into account, which recently
have been shown to be more efficient than previously estimated (\cf
Weinberg 1994; Kundi\'c \& Ostriker 1995).

\begin{figure}
\begin{center}
\leavevmode
\epsfxsize 11cm
%\epsffile{iau174figs/piet_fig3.ps}
\epsffile{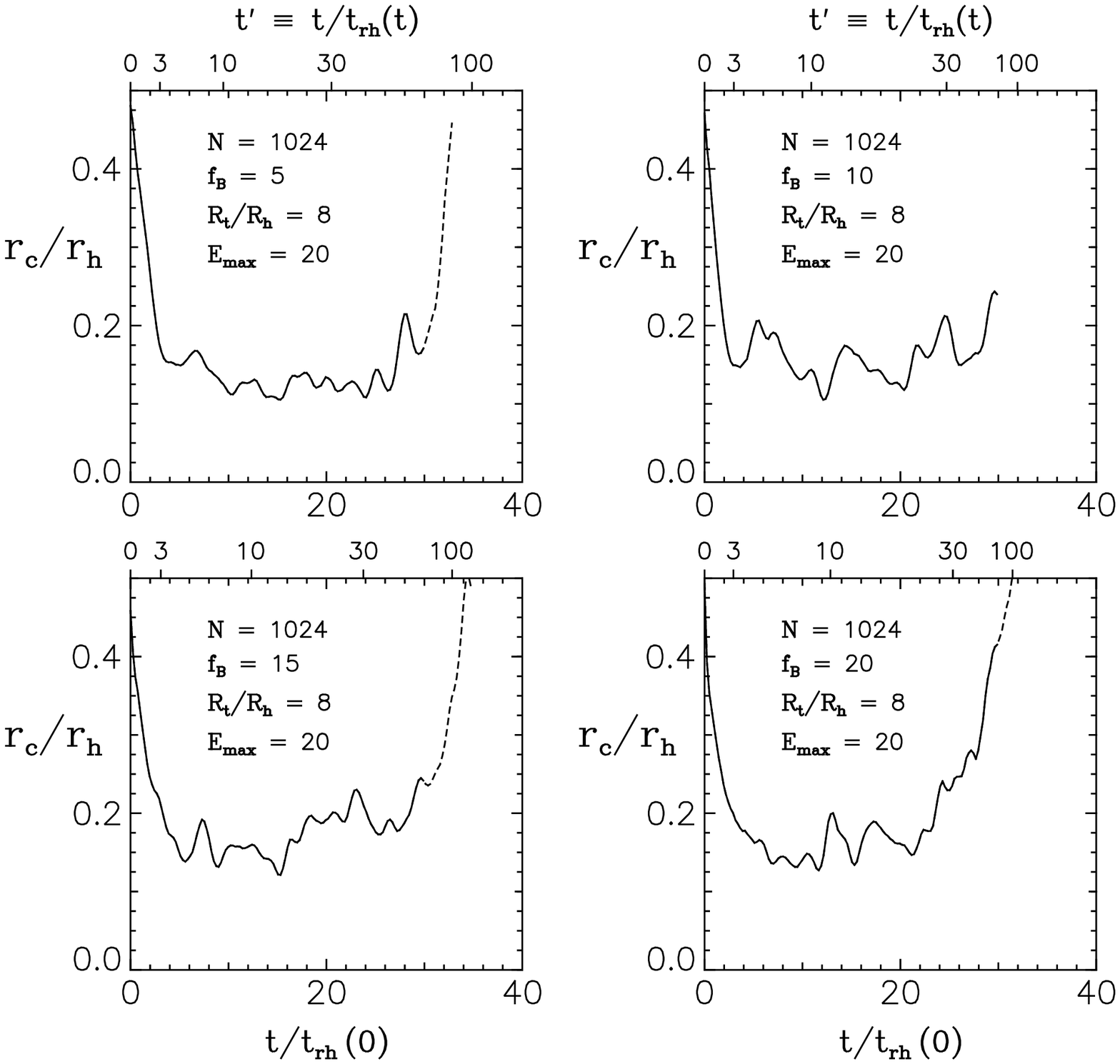}
\caption{Core radius $r_c$ in units of the half mass radius $r_h$, as
a function of time; labels as in fig. 1.}
\end{center}
\end{figure} 

In fig. 2, the core population of binaries is followed.  Plotted is
\fbcc, the fraction of objects in the core that are binaries, rather
than single stars or multiple star systems containing more than two
stars.  Again, it is clear that for $f_B < 15\%$ the core binary
population will eventually drop to very low values, whereas for
$f_B > 15\%$ the core binary population will remain constant,
within the numerical noise.  The fraction of the cluster mass that is
present in the core will increase towards the end of the lifetime of
the cluster, as the outer regions are eaten away by the galactic tidal
field (see McMillan and Hut 1994 for a detailed discussion).  For
values of \fb\ above the watershed, this implies a growing core size,
in units of the half mass radius, as is evident in Fig. 3.  For
smaller values of \fb, however, there is a near-cancellation of two
tendencies, resulting in a near-constant value for $r_c/r_h$.  An
isolated cluster would show a shrinking core (fewer binaries lead to a
higher density required for the same energy generation rate, fixed by
the rate of energy loss at this outskirts), and an expanding half-mass
radius.  Both effects would lead to a decrease in $r_c/r_h$.  A
tidally truncated cluster, however, would show a decrease in $r_h$ as
well as a decrease in $r_c$.  In the case of the present simulations,
these two effects turn out to be comparable in magnitude.

\section{Summary and Discussion}

It is quite possible that primordial binaries may provide the only
type of fuel needed to power post-collapse evolution, for most
globular clusters (McMillan \& Hut 1994).  In this case deep core
collapse will not occur, and the core will contain a significant
fraction of binaries at any time, even though the rest of the cluster
may become more depleted in binaries through mass segregation.

For those clusters that do run out of primordial binaries, subsequent
post-collapse evolution most likely will show chaotic core
oscillations, with relatively large core radii almost all of the time,
comparable to the case in which primordial binaries are still present.
In both cases, $r_c/r_h \sim 10^{-2}$, within a factor of a few.  The
precise value of this factor is crucial for comparison with
observations, but has not been determined so far from simulations.
Fokker-Plank methods are inherently not well fit to deal with
primordial binaries, and direct $N$-body calculations with primordial
binaries have not yet been carried out for sufficiently large $N$
values.  Fortunately, the GRAPE-4 computer will be up to the
challenge, and is expected to provide us with the answer within a year
or so.  At that time, we will be able to judge whether the result
reported by Guhathakurta (these proceedings), that in M15 the core
radius $r_c < 0.1$pc, is compatible with either or both scenarios
(primordial binary burning versus gravothermal oscillations).

\section*{Acknowledgements}

I thank Steve McMillan for his help in constructing the figures, based
upon the runs reported in McMillan and Hut (1994).  I thank Profs.
Sugimoto and Makino, as well as the whole GRAPE group, for their
hospitality during my visit to Tokyo University, when this paper was
written.  I am also grateful for the grant from JSPS which enabled me
to make this visit.

\end{document}